\documentclass[11pt]{article}
\usepackage{jheppubmodessay}
\pdfoutput=1
\usepackage[showonlyrefs]{mathtools}
\usepackage{psfrag}
\usepackage{array}
\usepackage{amssymb, amsthm}
\usepackage{graphicx}
\newtheorem*{conjecture}{Conjecture}
\usepackage[labelsep=quad]{subcaption}
\mathtoolsset{showonlyrefs}
\usepackage{epstopdf}
	
\usepackage{epsfig}
\usepackage[punctsep]{collref}
\collectsep[]{;}	
\def\projvac{{\mathcal P}_{\Omega}}
\def\alcan{{\mathcal A}_{\text{bdry}}}
\def\hcan{H}

\def\Or[#1]{{\text{O}}\left({#1}\right)}
\def\dotl[#1,#2]{\left\langle #1,\, #2 \right\rangle}
\def\dotlb[#1,#2]{\left\langle #1,\, #2 \right\rangle}
\def\dotlm[#1,#2]{\left[ #1,\, #2 \right]}
\def\dotp[#1,#2]{(\vect{#1} \cdot\vect{#2})}
\def\aff[#1,#2]{\hat{#1}(#2)}
\def\n4sym{{\cal N}=4 SYM}
\def\>{\rangle}
\def\<{\langle}
\def\weight[#1,#2,#3]{\{(#1),#2,#3\}}
\def\ads[#1]{$\text{AdS}_{#1}$}
\def\cft[#1]{$\text{CFT}_{#1}$}

\newcommand{\be}{\begin{equation}}
\newcommand{\ee}{\end{equation}}
\newcommand{\ba}{\begin{align}}
\newcommand{\ea}{\end{align}}
\newcommand{\bs}{\begin{split}}
\def\sess\end{split}

\newcommand{\vect}[1]{{\boldsymbol{#1}}}

\notoc
\title{Is Holography Implicit in Canonical Gravity?}
\author{Suvrat Raju}
\affiliation{International Centre for Theoretical Sciences, Tata Institute of Fundamental Research, Shivakote, Bengaluru 560089, India.}
\emailAdd{suvrat@icts.res.in}
\date{}

\abstract{We conjecture that, in asymptotically anti-de Sitter space, two solutions of the Wheeler-DeWitt equation that coincide asymptotically must also coincide in the bulk. This suggests that the essential elements of holography are already present in canonical gravity.  Our argument sheds light on why holography works only in gravitational theories, and also on the significance of anti-de Sitter boundary conditions.
\vskip 1in
{\bf {\center{Essay written for the Gravity Research Foundation 2019 Awards for Essays on Gravitation.}}}
}

\begin{document}
\maketitle
\section{Introduction}
The discovery of holography \cite{Maldacena:1997re} --- the idea that gravitational dynamics in $d+1$ dimensions may be described by a $d$-dimensional non-gravitational theory --- is an extremely important development in quantum gravity.  However, holography is sometimes believed to be a mysterious phenomenon that emerges from the intricacies of string theory. In this essay, we will discuss the perhaps surprising idea that the elements of holography are already present in the canonical formulation of gravity when we consider spacetimes with asymptotically anti-de Sitter (AdS) boundary conditions.  Similar ideas were explored in \cite{Marolf:2008mf} although our perspective will be different and will follow \cite{Banerjee:2016mhh,Raju:2018zpn}.

To be clear: the claim in this essay is not that the canonical description of pure gravity, by itself, leads to a complete theory. Rather, we would like to suggest that any complete theory of quantum gravity, which subsumes the formalism of canonical gravity within itself,  must display holography in a sense that we make precise. Furthermore, this inference does not require any detailed knowledge about ultraviolet (UV) physics but merely a careful analysis of low-energy physics that we already understand.

We start by reviewing some elementary facts about AdS quantum gravity.  Then we formulate the question about canonical holography in the form of a sharp conjecture about wavefunctions that satisfy the Wheeler-DeWitt (WdW) equation:  in a theory of gravity with asymptotically AdS boundary conditions, all the information in this wavefunction can be extracted purely from asymptotic data at one point of time. After laying out some physical assumptions, we provide evidence for this conjecture. We conclude with a discussion of some open questions.

\section{Quantum gravity in anti-de Sitter space}
We will consider a theory of gravity in $d+1$ spacetime dimensions, possibly coupled to matter, and with a negative cosmological constant, $\Lambda$. We choose units so that  $2 \Lambda = -{d (d - 1)}$.

\begin{figure}[!h]
\begin{center}
\includegraphics[height=0.3\textheight]{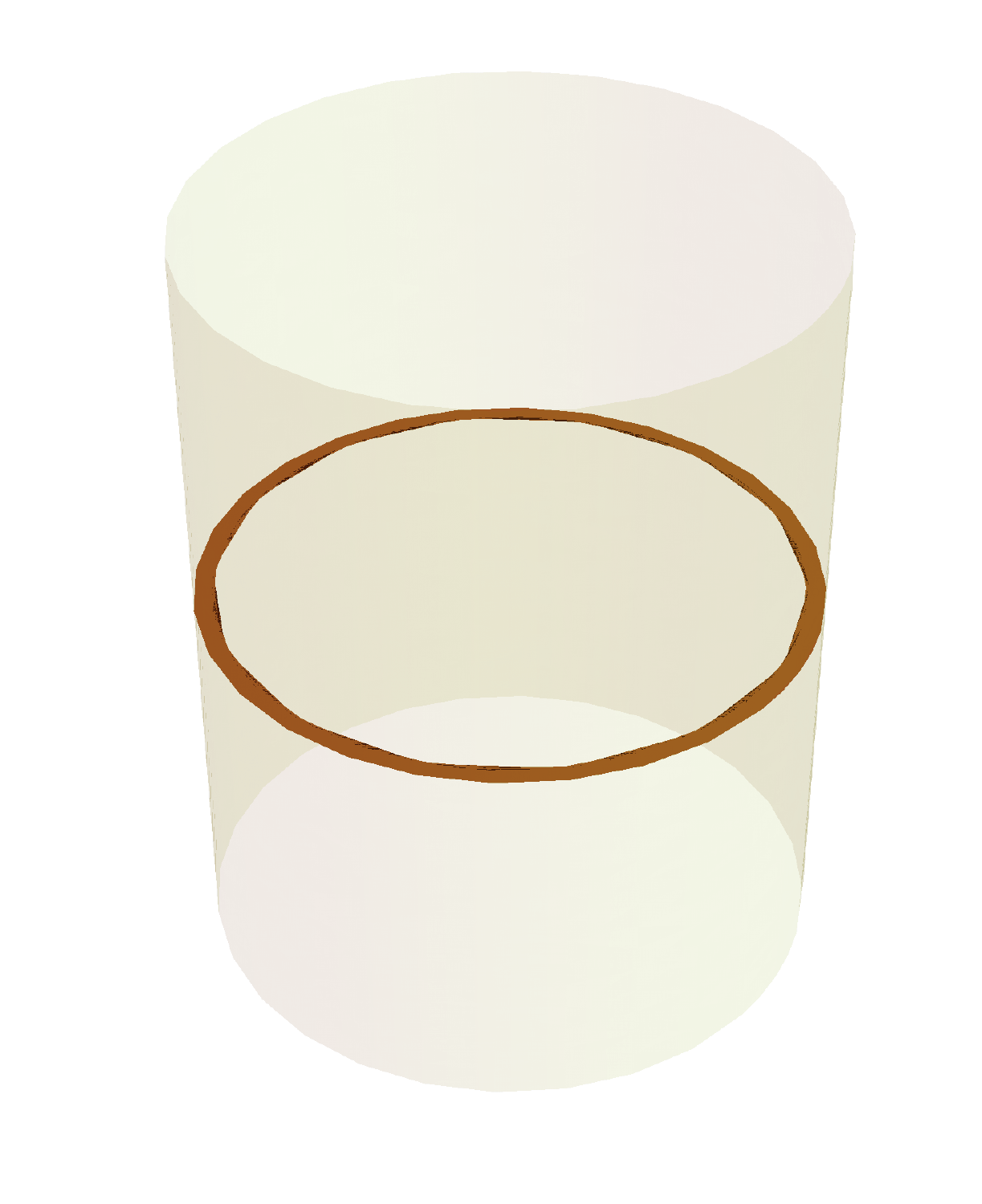}
\end{center}
\caption{\em AdS can be visualized as a solid cylinder.  The asymptotic region, at $t = 0$,  is shaded in brown. \label{figads}}
\end{figure}

The central object in canonical gravity is the wavefunction over $d$-geometries. In theories with matter, the wavefunction also specifies the amplitudes for the matter fields to take on various configurations. We are interested in wavefunctions that have support exclusively on those metrics that are asymptotically AdS.   To set notation, we remind the reader that the metric of the pure AdS spacetime is
\be
ds^2 = g^{\text{AdS}}_{\mu \nu} dx^{\mu} dx^{\nu} = {-(r^2 + 1) dt^2}  + {dr^2 \over r^2 + 1} + r^2 d \Omega_{d-1}^2,
\ee
where the AdS radius is set to unity by the choice of $\Lambda$ above. The spacetime described by this metric can be visualized as in Figure \ref{figads} and its boundary is $S^{d-1} \times R$.

It is important to recognize that although the leading asymptotic behaviour of the metric is fixed, we still allow {\em normalizable} fluctuations of the metric and other matter fields near the AdS boundary \cite{Henneaux:1985tv}. If we expand the field as $g_{\mu \nu} = g_{\mu \nu}^{\text{AdS}} + h_{\mu \nu}$, then the form of the allowed fluctuations can be specified most easily by choosing the gauge $h_{r \mu} = 0$ near the boundary. For the other components, $h_{i j}$,  we demand that, as $r \rightarrow \infty$, $h_{i j}$ fall  off at least as fast as $r^{2 - d}$. If the theory contains a massive scalar field, $\phi$,  with non-negative $m^2$, then we allow fluctuations of this field that fall off as fast as $r^{-\Delta}$ where $2 \Delta = d + (d^2 + 4 m^2)^{1 \over 2}$. 

Therefore, even in the canonical theory, the asymptotic values of these fluctuations, such as
\be
\label{asymptop}
t_{i j} = \lim_{r \rightarrow \infty} r^{d-2} h_{i j}; \quad \text{and} \quad O = \lim_{r \rightarrow \infty} r^{\Delta} \phi,
\ee
and similar operators for other fields in the theory,  are good observables.  The canonical wavefunction allows us to compute quantum correlation functions of products of such operators. 

We will work in the Schrodinger picture, and so  the reader can think of the operators above as being defined at $t=0$. The operators also vary on $S^{d-1}$ but we have suppressed that dependence to lighten the notation. By considering arbitrary bounded functions of these elementary asymptotic operators, we obtain a von Neumann algebra, which we will call $\alcan$.

\section{A conjecture about canonical holography}
In canonical gravity, the allowed wavefunctions must satisfy the WdW  equation. The WdW equation imposes the invariance of the wavefunctions under  ``small diffeomorphisms'' --- those that die off at the boundary of AdS. However, the wavefunctions are allowed to transform under ``large diffeomorphisms'' --- those that do not vanish at the boundary.  So gravitational physics in asymptotically AdS spacetimes is not ``timeless'' and we can label the wavefunctions unambiguously by the boundary time, $t$. 

The claim that canonical gravity is holographic can then be stated in the form of the following conjecture.
\begin{conjecture}
   In asymptotically anti-de Sitter space, if two solutions of the WdW equation agree exactly at the boundary at $t = 0$, then they are equal.
\end{conjecture}

The statement that the wavefunctions agree asymptotically means that they give exactly the same expectation value to any element of $\alcan$.  We emphasize that we impose the equality of these expectation values  only at $t = 0$ and {\em not} for all time. More visually, we demand that the wavefunctions coincide on the brown strip shown in Figure \ref{figads}. The conjecture is that imposing this constraint is sufficient to ensure that the wavefunctions yield the same expectation value for all operators in the theory i.e. they cannot differ even for bulk operators.

We immediately note that such a conjecture could only be true in the quantum theory. In the classical theory, it is clearly possible to find metrics and field configurations that coincide exactly in their asymptotics at $t = 0$ but differ significantly in the bulk.  Nevertheless, we will now provide some evidence that the quantum theory has this remarkable property. 

\section{Some evidence for canonical holography}
A striking fact about canonical gravity is that the Hamiltonian is weakly equal to a {\em boundary term} \cite{Regge:1974zd}. In AdS, a simple explicit expression is provided by \cite{deHaro:2000vlm}
\be
\label{hcanbdry}
\hcan =    {d \over 16 \pi G_N} \int  d^{d-1} \Omega \, t_{t t}.
\ee
This Hamiltonian  gives the energy of a state {\em above} the energy of global anti-de Sitter space when evaluated on wavefunctions that satisfy the WdW equation. 

We will need to  assume that when  the canonical theory is consistently UV-completed, the Hamiltonian can still be evaluated through a boundary term.  This is a good assumption since it just relies on the fact that the quantum wavefunction is invariant under bulk diffeomorphisms. Presumably, any complete theory of quantum gravity will respect this basic property. Moreover, in controlled examples of holography, provided by the AdS/CFT conjecture,  the relation between asymptotic fluctuations of the metric and the Hamiltonian is just an element of the so-called ``extrapolate'' dictionary \cite{Banks:1998dd}.

The low-energy spectrum of $\hcan$  is quite simple. First, it has a unique ground state  ---  empty global AdS --- which we denote by $|\Omega \rangle$, and whose eigenvalue is $0$.\footnote{The careful reader will note that
 in \eqref{hcanbdry} we have subtracted off the constant given in \cite{deHaro:2000vlm} to ensure $\hcan | \Omega \rangle = 0$.  If  \eqref{hcanbdry} is corrected at higher orders in $G_N$ we will assume for convenience that, if necessary, the redefinition  $\hcan \rightarrow \hcan - \langle \Omega | \hcan | \Omega \rangle$ has been carried out so that this remains the case.}    The lowest excitations above this ground state are separated by an energy gap.  The precise value of the gap depends on the field-content but, for massless particles, the lowest allowed value of the energy is $d$.  We will need to assume that, in the full quantum theory, this low-energy structure is not modified drastically i.e. that the full Hamiltonian continues to have a unique ground state, with well-separated excited states. This is also a good assumption: it only restates the expectation that whatever UV-physics is required to complete the theory does not change low-energy physics.

We now turn to our argument. Say that we have two distinct solutions of the WdW equation, which we denote by $|\Psi_1 \rangle$ and $|\Psi_2 \rangle$, that nevertheless give the same expectation value to all elements of $\alcan$.  Since the states are distinct, there must be {\em some} operator, $Q$,  that distinguishes between them.  Once we have quantized the theory to obtain a basis for the Hilbert space, whose elements we denote by $|n \rangle$, we can expand $Q$ in this basis: $Q = \sum_{n, m} c_{nm} |n \rangle \langle m|$. 
Now, consider the operator 
\be
P(z) =  e^{-z \hcan}.
\ee
It is clear that $P(z) \in \alcan$.  Moreover, since by our assumption,  $\hcan$ has a unique ground state that is separated from the nearest excited state by a gap, for values of $z$ larger than the inverse of this gap, the action of $P(z)$ tends to annihilate any excited state. Therefore, in the limit of large $z$, $P(z)$ just projects onto empty global AdS.
\be
\projvac \equiv \lim_{z \rightarrow \infty} P(z) =  |\Omega \rangle \langle \Omega |.
\ee

We now state a third assumption: it should be possible to formulate the theory within the span of states obtained by starting with the ground state in the far past,  and then  injecting energy by acting with asymptotic operators at various value of time.  This assumption states  that the Hilbert space is  spanned by states of the form
\be
\label{setstates}
e^{-i H t_n} O_n \ldots e^{-i H t_2} O_2 e^{-i H t_1} O_1 |\Omega \rangle,
\ee
where $O_i \in \alcan$ and $t_i$ are arbitrary real numbers.
This is a good assumption since the set of states evidently displayed in \eqref{setstates} is manifestly closed under dynamical evolution. So, considerations of unitarity, by themselves, cannot force us to enlarge the Hilbert state. As a consistency check, at low-energies, it is easy to check explicitly that states in the Hilbert space of bulk effective field theory can all be written in the form \eqref{setstates}. The set \eqref{setstates} is also rich enough to describe all black holes that can be formed through collapse of matter injected from the boundary.\footnote{It is possible that \eqref{setstates}  excludes the so-called ``bags of gold'' states, which are solutions where the black-hole interior contains an entire Friedmann-Robertson-Walker universe. But, if so, since \eqref{setstates} is closed under time-evolution, this implies that the ``bags of gold'' form a separate superselection sector of states that cannot evolve to states of the form \eqref{setstates} or be obtained by collapse from states of the form \eqref{setstates} \cite{Marolf:2008tx}.}

The parameterization used in \eqref{setstates} has the simple physical interpretation described above but it is redundant. Since the spectrum of the Hamiltonian is positive, the inner-product of the vector displayed in \eqref{setstates} with any other vector is analytic when the $t_i$ are extended in the lower-half complex plane. Then the ``edge of the wedge'' theorem can be used to show that the Hilbert space can be densely populated by acting with operators just in an infinitesimal time strip i.e. by taking $t_i \in [0, \epsilon]$ for any finite $\epsilon$  in \eqref{setstates}.  But even this infinitesimal strip is redundant, and we can use our first assumption to squeeze it down to an instant. This is because, since $H \in \alcan$,  the entire product of operators displayed in \eqref{setstates} can be represented as a single element of $\alcan$. Therefore our assumptions imply that the {\em  action of $\alcan$ on the vacuum generates the entire Hilbert space.} In algebraic quantum field theory language \cite{Haag:1992hx}, this is the conclusion that the vacuum is {\em cyclic} with respect to the action of $\alcan$.

But then, {\em each} element of the basis introduced above can be written as
\be
\label{reehschlieder}
|n \rangle = X_n |\Omega \rangle.
\ee
Here $X_n$ is an {\em asymptotic} operator and an element of $\alcan$, although it might be a very complicated function of the elementary asymptotic operators displayed in \eqref{asymptop}.

This implies
\be
Q = \sum_{n,m} c_{n m} X_n \projvac X_m^{\dagger}.
\ee
However, the right-hand side above {\em only} makes reference to asymptotic operators. Therefore, by assumption,  the two wavefunctions, $|\Psi_1 \rangle$ and $|\Psi_2 \rangle$ must give identical expectation values to the right hand side. But this contradicts the initial assumption that they give different expectation values to $Q$. Therefore the operator $Q$ cannot exist. 

This establishes the conjecture subject to the three physical assumptions made above --- first, that the  Hamiltonian can be evaluated through a boundary term, second, that its low-energy spectrum is not changed by UV-physics, and, third, that we can consistently formulate the theory within the set of states obtained by exciting the ground state with asymptotic operators.

\section{Discussion}
The reader may be concerned that the argument above was too quick, and so we explore some of its physical foundations.

\paragraph{\bf The quantum nature of holography.} A surprising aspect of the conjecture above is that it has no classical analogue. However, it is not difficult to see, even without a detailed argument, that asymptotic observables
carry more information in the quantum theory than in the classical theory. In the quantum theory,  not only can we measure $\langle \hcan \rangle$ in any state asymptotically, we can, in principle, measure all
the higher moments, $\langle \hcan^q \rangle$. There is no analogue of this data in the classical theory. If the state is expressed as a linear combination of energy eigenstates with complex coefficients, these moments give us information about the {\em magnitudes} of all these coefficients just from asymptotic observations. These measurements, by themselves,
do not help us determine the phases of these coefficients or identify the state uniquely if the energy-spectrum is degenerate; this explains the need for the operators $X_n$ in the argument above.

\paragraph{\bf  Gravity vs other gauge theories.} A second question that may occur to the reader is that, in any gauge theory, the charge can be measured through a boundary term. Since an important ingredient in the argument above was that  the Hamiltonian could be evaluated through a boundary term, could one make a similar conjecture in a gauge theory without gravity? 

However, a little thought shows that non-gravitational gauge theories {\em cannot} be holographic.  Let $|\Psi_1 \rangle$ be any state and take $|\Psi_2 \rangle = U |\Psi_1 \rangle$ where $U$ is a small Wilson loop operator that acts in the bulk at $t = 0$. This Wilson-loop operator is spacelike to the asymptotic region at $t = 0$, and since it is gauge-invariant, it commutes with all observables near the boundary at $t = 0$ in a non-gravitational theory. Therefore {\em no} asymptotic operator can distinguish between $|\Psi_1 \rangle$ and $|\Psi_2 \rangle$ although the states are clearly different. 

The reason that asymptotic observations cannot uniquely identify the state in a gauge theory is that,  for any given charge (including zero charge), there is an {\em infinite} degeneracy of states. In contrast, there is a unique ground state in gravity with asymptotically AdS boundary conditions and we can exploit this to read off the bulk state from asymptotic observables.

\paragraph{\bf Holography in flat space?}
Could one extend the argument presented here to the case with vanishing cosmological constant? At first sight, the subtlety is as follows.  The construction of $\projvac$ assumed the existence of a unique vacuum, with clearly separated excited states.    The infra-red structure of flat space is different, and so a similar construction would project onto a family of states distinguished only by their soft content. However, this does not seem to be an insuperable obstacle, since it should be possible to obtain a finer projector than $\projvac$ by using the full set of asymptotic symmetries \cite{Strominger:2017zoo}. This is a very interesting direction for future work.

\paragraph*{Acknowledgments.} I am grateful to Rajesh Gopakumar, R. Loganayagam, Kyriakos Papadodimas,  Madhavan Varadarajan and Spenta Wadia for useful discussions. This work was partially supported by a Swarnajayanti fellowship of the Department of Science and Technology (India).

\bibliographystyle{JHEP}
\bibliography{references}

\end{document}